\documentclass{iau} 
\usepackage{graphicx}

\title[Long-term photometric observations in the field of the star formation region NGC7129] 
{Long-term photometric observations in the field of the star formation region NGC7129}

\author[Evgeni Semkov, Stoyanka Peneva, Sunay Ibryamov \& Asen Mutafov]   
{Evgeni Semkov$^1$, Stoyanka Peneva$^1$, Sunay Ibryamov$^1$$^,$$^2$
 \and Asen Mutafov$^1$}

\affiliation{$^1$Institute of Astronomy and National Astronomical Observatory, Bulgarian Academy of Sciences, Sofia, Bulgaria 
\\ email: {\tt esemkov@astro.bas.bg} \\[\affilskip]
$^2$Department of Theoretical and Applied Physics, University of Shumen, Shumen, Bulgaria}

\pubyear{2017}
\volume{336}  
\jname{Astrophysical Masers: Unlocking the Mysteries of the Universe}
\editors{A. Tarchi, M.J. Reid \& P. Castangia, eds.}
\begin{document}

\maketitle

\begin{abstract}
We present results from long-term optical photometric observations of the Pre-Main Sequence (PMS) stars, located in the star formation region around the bright nebula NGC 7129. 
Using the long-term light curves and spectroscopic data, we tried to classify the PMS objects in the field and to define the reasons for the observed brightness variations.
Our main goal is to explore the known PMS stars and discover new, young, variable stars.
The new variable PMS star 2MASS J21403576+6635000 exhibits unusual brightness variations for very short time intervals (few minutes or hours) with comparatively large amplitudes ($\Delta I$ = 2.65 mag).

\end{abstract}

\firstsection 
\section{Introduction}

The region NGC 7129 is a part of a larger structure, called Cepheus Bubble, and its represents a region with active star formation. The presence of a large number of Herbig-Haro objects, collimated jets, Herbig's Ae/Be and T Tauri stars, water masers, molecular outflow and other young objects have been reported in previous studies in the region (\cite[Hartigan \& Lada 1985]{hartigan85}; \cite[Miranda \etal\ 1995]{miranda93}, \cite[Magakian \& Movsessian 1997]{magakian97}; \cite[Kun \etal\ 2009]{kun09}). Using recent data from photometric monitoring and data from the photographic plate archives we aim at studing, the long-term photometric behavior of PMS objects in the field. 

Our main goal is to study in more detail the known PMS objects in the field of NGC 7129 and to search for new, young, variable stars.
The variability of PMS stars manifests itself as transient increases or temporary drops in brightness or as large amplitude irregular or regular variations. 
During our photometric monitoring, several PMS stars were found that could be classified as T Tauri stars (\cite[Semkov 2006]{semkov06}).
Such a long-term study of the fields of star formation has shown its effectiveness in discovering new phenomena important for astronomy (\cite[Semkov \etal\ 2010]{Semkov2010}).

\begin{figure}[b]
\begin{center}
 \includegraphics[width=12cm]{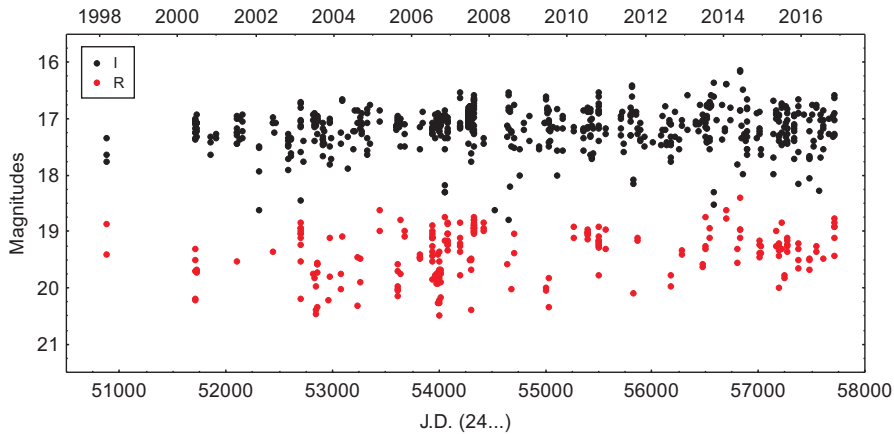} 
 \caption{$RI$ light curves of V4 for the period February 1998$-$November 2016.}
   \label{fig1}
\end{center}
\end{figure}

\section{Results and Discussion}

Our optical photometric observations were performed with the 2-m RCC, 50/70-cm Schmidt and 60-cm Cassegrain telescopes of the National Astronomical Observatory Rozhen (Bulgaria) and the 1.3-m RC telescope of the Skinakas Observatory (Crete, Greece). The observations were performed with eight different types of CCD cameras. The technical parameters for the cameras used, observational procedure and data reduction process are described in \cite[Ibryamov \etal\ (2015)]{ibryamov15}. Most suitable for long-term photometric study are the plate archives of the big Schmidt telescopes that have a large field of view, as the 105/150 cm Schmidt telescope at Kiso Observatory, the 67/92 cm Schmidt telescope at Asiago Observatory, the Palomar Schmidt telescope and others (\cite[Semkov \etal\ 2013]{semkov13}).

The two most extensively studied PMS objects in the field of NGC 7129 are the stars V350 Cep and V391 Cep.
Both stars show strong photometric variability and a spectrum rich of emission lines typical of classical T Tauri stars.
The historical $B$-light curve of V350 Cep is very similar to eruptive PMS stars from FU Orionis type, but its spectrum is similar to another type of eruptive stars - EX Lupi.
It is very likely that V350 Cep is an intermediate object between the two types of PMS stars showing large amplitude outbursts.
A characteristic features of V391 Cep are periods of high amplitude brightness variability, followed by periods of smaller amplitude variability.

The star 2MASS J21403576+6635000 (hereafter V4) was recently discovered as a PMS variable. 
V4 shows very strong and fast photometric variability during short time periods (several minutes or hours) with large amplitude.
The brightness of V4 during the period of our photometric observations 1998$-$2016 varies in the range 16.14$-$18.79 mag for the $I$-band and 18.40$-$20.48 mag for the $R$-band. 
Evidences of periodicity in the brightness variability of V4 are not detected.
It can be seen from Fig. 1 that during our study the brightness of V4 vary around some intermediate level. 
The presence of V4 in the field of star formation and the irregular variability with large amplitude suggest a PMS nature of the star.  
We suggest that the V4 is a T Tauri star and the observed fast variability with large amplitudes probably is caused from a strong irregular accretion rate from circumstellar disk onto the stellar surface.

\end{document}